\documentclass[useAMS]{mn2e}

\begin{document}

\title[The Photometric Plane]
      {The Photometric Plane of Elliptical Galaxies} 
\author[Alister W.~Graham]
       {Alister W.~Graham\thanks{Present Address: Department of Astronomy, 
  University of Florida, Gainesville, FL, USA}\\
  Instituto de Astrof\'{i}sica de Canarias, La Laguna, E-38200, Tenerife.
        Spain}
\date{Received 2002 January 10; Accepted 2002 March 28}

\pubyear{2002} \volume{000} 
\pagerange{\pageref{firstpage}--\pageref{lastpage}}

\input{psfig}

\maketitle
\label{firstpage}

\begin{abstract}

The S\'ersic ($r^{1/n}$) index $n$ of an elliptical galaxy (or bulge) 
has recently been shown to correlate strongly ($r=0.8$) with a 
galaxy's central velocity dispersion.  
This index could therefore prove extremely useful and 
cost-effective (in terms of both telescope time and data reduction) 
for many fields of extragalactic research. 
It is a purely photometric quantity which apparently not only traces 
the mass of a bulge but has additionally been shown to reflect the 
degree of bulge concentration.  
This paper explores the affect of replacing the central velocity 
dispersion term in the Fundamental Plane with the S\'ersic index $n$. 
Using a sample of early-type galaxies from the Virgo and Fornax clusters, various 
($B$-band) `Photometric Planes' were constructed and found to have a scatter 
of 0.14-0.17 dex in $\log r_{\rm e}$, or a distance error of 
38-48 per cent per galaxy (the higher values arising from the inclusion
of the S0 galaxies).  The corresponding Fundamental Plane yielded 
a 33-37 per cent error in distance for the same galaxy (sub-)samples 
(i.e.\ $\sim$15-30 per cent less scatter).  
The gains in using a hyperplane (i.e.\ adding the S\'ersic index 
to the Fundamental Plane as a fourth parameter) 
were small, giving a 27-33 per cent error in distance, 
depending on the galaxy sample used. 
The Photometric Plane has been used here to estimate the 
Virgo-Fornax distance 
modulus; giving a value of $\Delta\mu=0.62\pm0.30$ mag (cf.\ 
0.51$\pm$0.21, HST Key Project on the Extragalactic distance Scale). 
The prospects for using the Photometric Plane at higher redshift
appears promising.  Using published data on the intermediate redshift 
cluster Cl 1358+62 ($z$=0.33) gave a Photometric Plane distance 
error of 35-41 per cent per galaxy. 

\end{abstract}

\begin{keywords}
distance scale -- galaxies: elliptical and lenticular, cD -- galaxies: fundamental parameters -- galaxies: kinematics and dynamics -- galaxies: photometry -- galaxies: structure
\end{keywords}

\section{Introduction}

Studies of supermassive black holes and the elliptical galaxies
(and bulges) at whose centers they reside have revealed that the 
S\'ersic (1968) index $n$ can be used just as effectively for 
predicting the mass of the black hole as the 
velocity dispersion of the stars (Graham et al.\ 2001).  
Aside from the physical 
insight this provides into the formation of galaxy and black 
hole alike, it also has important practical consequences: relatively
expensive spectroscopic observations can be replaced with 
(photometrically uncalibrated) images.  
Furthermore, the issue of aperture corrections for velocity 
dispersion measurements is bypassed completely. 

In this paper we explore 
whether or not the `Fundamental Plane' (Djorgovski \& Davis 1987;
Dressler et al.\ 1987) remains as thin (or 
at an acceptable/useful thickness) when the central velocity 
dispersion term is replaced with the S\'ersic index $n$, producing 
what is called a `Photometric Plane'.


Khosroshahi et al.\ (2000a) constructed a near-infrared 
Photometric Plane from a sample of 42 Coma elliptical 
galaxies and the bulges of 26 early-type disk galaxies taken from 
the field (Khosroshahi, Wadadekar \& Kembhavi 2000b). 
Their Photometric Plane was constructed using the S\'ersic 
index $n$, the effective radius 
$r_{\rm e}$ and the central bulge surface brightness 
$\mu_0$ derived from the best-fitting $r^{1/n}$ model. 
The authors claimed that the scatter about their (elliptical only) 
Photometric Plane implied an error of 53 per cent in the derived 
distance to any single galaxy. 
%
%
%
%
M\"ollenhoff \& Heidt (2001) modelled 40 early-type 
spiral galaxy bulges and computed a correlation coefficient $r=0.91$ 
between $\log n$ and a linear combination of $\mu_0$ and $\log r_e$. 
Although they did not give an estimate to the scatter, a strong
correlation clearly exists. 

In the present analysis, the Photometric Plane will be derived using 
a well known sample of Virgo and Fornax elliptical galaxies for which 
the required data has already been published. 
Importantly, both the Photometric Plane {\it and} 
the Fundamental Plane will be derived for exactly the same galaxy sample, 
enabling a direct comparison of the scatter about the two planes.  
The usefulness of the Photometric Plane will subsequently be tested by 
computing the Virgo-Fornax distance modulus and comparing the 
result with the latest estimate from the {\it HST Key Project on the 
Extragalactic Distance Scale}.

Section 2 introduces the data which has been collated from the
literature.  The construction and analysis of the Photometric
and Fundamental Planes for the full galaxy sample (and various 
sub-samples) is presented in Section 3.  A hyperplane using all 
three S\'ersic (photometric) parameters plus the central velocity 
dispersion is also introduced here.  The analysis is 
discussed in Section 4, and a brief comparison and/or discussion 
is made of several other purely photometric relations such as the 
Kormendy relation, the scalelength-shape ($r-n$) relation of Young 
\& Curry (1994) and the `Entropic Plane' (Lima Neto et al.\ 1999).  
Lastly, the Photometric Plane of the intermediate redshift 
cluster Cl 1358+62 ($z$=0.33) is constructed and shown to have 
comprable scatter to the local Photometric Plane.

\section{Kinematic and Photometric Data}

The photometric data for the present investigation have been taken
from table 2 of Caon, Capaccioli \& D'Onofrio (1993) and  
table A1 of D'Onofrio, Capaccioli \& Caon (1994).   
The data is discussed further in Caon, Capaccioli \& Rampazzo (1990) 
and Caon, Capaccioli \& D'Onofrio (1994a).  
In essence, it consists of a ($B$-band) magnitude-limited sample of 
elliptical and non-barred S0 galaxies which is 
100 per cent complete down to $B_T$=15 mag for the Fornax cluster and 
80 per cent complete down to $B_T$=14 mag for the Virgo cluster. 
These tables give 
the {\it model-independent} equivalent half-light radii 
($r_{\rm e}=\sqrt{a_{\rm e}b_{\rm e}}$, where $a_{\rm e}$ and 
$b_{\rm e}$ are the half-light radii of the semi-major and semi-minor
axis) 
%
%
and the $B$-band surface brightness at this radius ($\mu_{\rm e}$). 
These two quantities will therefore be used together with 
the {\it equivalent-profile} S\'ersic index $n_{\rm eq}$ ($n$ hereafter). 
The equivalent axis rather than the 
major-axis was selected because more galaxies have tabulated values of 
$n_{\rm eq}$ than $n_{\rm maj}$.   Typical erorrs for the value of $n$ 
are around 25\%, or 0.10 dex for $\log n$. 


No attempt has been made here to re-model the light profile data 
as this task was very recently performed by D'Onofrio (2001a, 2001b). 
Using his new data produced consistent results (see section 4) with 
the data referred to above.  

The trend of increasing central bulge intensity with increasing 
bulge luminosity (and $n$) which holds for spiral galaxy bulges 
(Khosroshahi et al.\ 2000b), and dwarf elliptical
and ordinary elliptical galaxies (Jerjen \& Binggeli 1997;
Graham, Trujillo \& Caon 2001) 
breaks down for the high-luminosity ellipticals  
(see, e.g.\ Faber et al.\ 1997, their figure 4).  
This is likely due, at least in part, to the presence of `cores' 
in bright elliptical galaxies (Ferrarese et al.\ 1994; Lauer et 
al.\ 1995).  Consequently, the (extrapolated) central surface 
brightness derived from a S\'ersic model may not always be 
realised, especially for those galaxies with large values of $n$.  
As the present investigation will use several bright 
elliptical galaxies, the central surface brightness (used by 
Khosroshahi et al.\ 2000a and M\"ollenhoff \& Heidt 2001) will
therefore not be used here.  Instead, the mean surface brightness 
within the effective half-light radius will be used.  
Values of $\mu_{\rm e}$ have been converted\footnote{See, e.g., 
Appendix A of Graham \& Colless 1997.} 
into $<\mu>_{\rm e}$, the mean surface brightness within $r_{\rm e}$, 
in units of mag arcsec$^{-2}$. 
This is what is typically used for constructing the Fundamental Plane.
When expressed in linear units, this term shall be denoted 
$\Sigma _{\rm e}$. 

Central velocity dispersion measurements ($\sigma_0$) have been 
taken from Hypercat\footnote{Hypercat can be reached at
http://www-obs.univ-lyon1.fr/hypercat/}. 
In order to make a direct comparison between the Fundamental Plane 
and the Photometric plane, only galaxies for which central velocity
dispersion measurements and $n_{\rm eq}$ are available have been used. 
This resulted in a final sample of 19 Es and 11 S0s from the Virgo 
cluster, and 8 Es from the Fornax cluster. 
This is therefore in no sense a statistically complete galaxy sample. 
The collated data set is given in Table~\ref{tab1}. 

\begin{figure}
\centerline{\psfig{figure=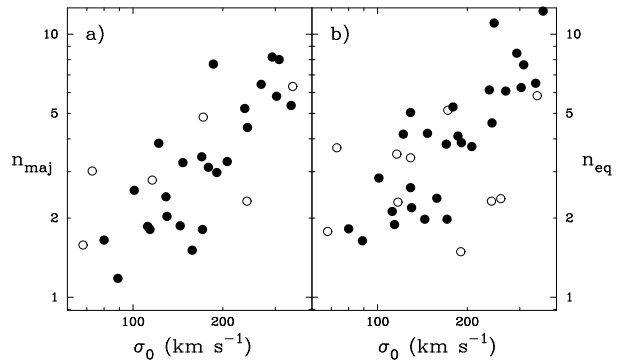,width=8.0cm,angle=-90}}
\caption{a) Major-axis and b) equivalent-axis S\'ersic index $n$ 
plotted against the central galaxy velocity dispersion.  
Elliptical galaxies are marked with filled circles, while S0 
galaxies are denoted by open circles.} 
\label{fig1}
\end{figure}

The central velocity dispersion has been plotted against the 
equivalent- (and major-) axis S\'ersic 
index in Figure~\ref{fig1}.  It is worth stating the obvious
here: these are two completely independently derived quantities. 
Despite this, and the heterogeneous nature of the velocity dispersion
data, for the elliptical galaxy sample the Spearman rank-order 
correlation coefficient is $r_s$=0.82 when using $n_{\rm eq}$ and 
$r_s$=0.83 when using $n_{\rm maj}$. 
Obviously $n$ is not simply some random third parameter in the 
$r^{1/n}$ model which produces better profile fits.  Quite the contrary, 
the S\'ersic index not only traces the mass of a galaxy (given that 
central velocity dispersion is a reliable estimate of mass) but is 
also known to quantify a galaxy's degree of concentration 
(Trujillo, Graham \& Caon 2001; Graham, Trujillo \& Caon 2001). 
The existence of colour gradients does however mean that 
$n$ will unfortunately be a function of bandpass, whereas stellar 
velocity dispersion should not.

\begin{table}
\centering
\caption{
The ($B$-band) photometric parameters given in columns 3-5 have 
been taken 
from table 2 and table A1 in Caon et al.\ (1993) and D'Onofrio 
et al.\ (1994) respectively, 
with the distinction that the mean surface brightness $<\mu>_{\rm e}$  
has been computed from the surface brightness $\mu_{\rm e}$ 
following Caon et al.\ (1994b). 
The value of $\log r_{\rm e}$ and $n$ correspond to that of the 
equivalent profile.  
Unlike the S\'ersic index $n$, the values of $r_{\rm e}$ and 
$\mu_{\rm e}$ were derived 
by the above authors independently of the S\'ersic $r^{1/n}$ model.
The central velocity dispersion (final column) has been taken
from Hypercat. 
\label{tab1}
}
\begin{tabular}{lccccc}
\hline
Galaxy & Morph.  &  $<\mu>_{\rm e}$  &  $\log r_{\rm e}$  &  $n$  &  $\sigma_0$ \\
Name   & Type    & mag arcsec$^{-2}$ & (kpc)  &       & km s$^{-1}$ \\
 \hline
N1339 &  E    &  20.46       &  0.07    &   1.98 &  171 \\
N1351 &  E    &  21.50       &  0.39    &   4.20 &  147 \\
N1374 &  E    &  20.94       &  0.35    &   3.74 &  207 \\
N1379 &  E    &  20.93       &  0.32    &   2.19 &  130 \\
N1399 &  E    &  22.16       &  1.05    &  12.24 &  359 \\
N1404 &  E    &  19.77       &  0.34    &   4.60 &  242 \\
N1419 &  E    &  20.20       & -0.08    &   5.04 &  129 \\
N1427 &  E    &  21.38       &  0.46    &   3.82 &  170 \\
N4168 &  E    &  21.57       &  0.44    &   4.10 &  186 \\
N4261 &  E    &  21.71       &  0.76    &   7.65 &  309 \\
N4365 &  E    &  22.12       &  0.93    &   6.08 &  269 \\
N4374 &  E    &  22.06       &  1.05    &   8.47 &  293 \\
N4387 &  E    &  20.59       &  0.03    &   2.12 &  112 \\
N4406 &  E    &  22.04       &  1.17    &  11.03 &  246 \\
N4434 &  E    &  20.59       &  0.06    &   4.17 &  122 \\
N4458 &  E    &  21.28       &  0.21    &   2.84 &  101 \\
N4464 &  E    &  19.72       & -0.20    &   2.61 &  129 \\
N4472 &  E    &  22.44       &  1.30    &   6.27 &  303 \\
N4473 &  E    &  20.66       &  0.48    &   5.29 &  179 \\
N4478 &  E    &  19.72       &  0.02    &   1.98 &  144 \\
N4486 &  E    &  21.88       &  1.05    &   6.51 &  339 \\
N4550 &  E    &  19.93       &  0.02    &   1.82 &   80 \\
N4551 &  E    &  20.67       &  0.06    &   1.89 &  114 \\
N4564 &  E    &  20.63       &  0.24    &   2.38 &  158 \\
N4621 &  E    &  22.39       &  0.96    &   6.14 &  237 \\
N4623 &  E    &  21.09       &  0.12    &   1.64 &   89 \\
N4660 &  E    &  19.53       &  0.06    &   3.87 &  191 \\
N4339 &  S0   &  21.50       &  0.37    &   3.50 &  116 \\
N4342 &  S0   &  18.48       & -0.36    &   2.32 &  241 \\
N4431 &  S0   &  22.65       &  0.28    &   1.78 &   68 \\
N4459 &  S0   &  20.64       &  0.44    &   5.14 &  172 \\
N4476 &  S0   &  21.03       &  0.13    &   3.70 &   73 \\
N4552 &  S0   &  22.24       &  0.95    &  12.81 &  263 \\
N4649 &  S0   &  21.63       &  0.96    &   5.84 &  343 \\
N4281 &  S0   &  21.10       &  0.34    &   2.37 &  259 \\
N4352 &  S0   &  21.78       &  0.20    &   2.30 &  117 \\
N4570 &  S0   &  19.92       &  0.15    &   1.49 &  190 \\
N4638 &  S0   &  19.55       &  0.09    &   3.39 &  129 \\
\hline 
\end{tabular}
\end{table}

\section{The Fundamental and Photometric Planes} 

\begin{figure}
\centerline{\psfig{figure=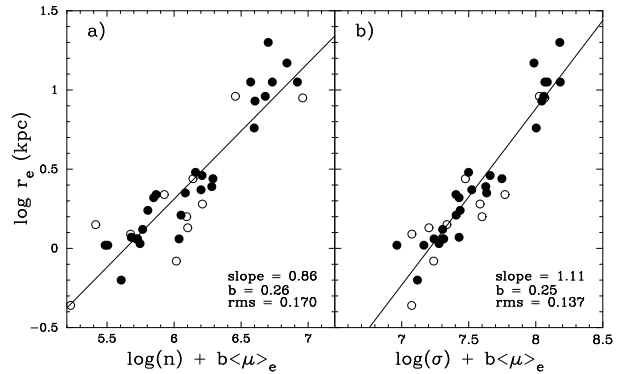,width=8.0cm,angle=-90}}
\caption{a) Photometric Plane. b) Fundamental Plane. 
Both planes have been constructed by minimising the vertical residual 
in $\log r_{\rm e}$. 
}
\label{fig2}
\end{figure}

Performing a least-squares regression analysis which minimises the 
scatter in the distance-dependent quantity $\log r_{\rm e}$ 
gave the Fundamental Plane relation
$r_{\rm e}\propto \sigma _{0}^{1.11\pm0.12}\Sigma _{\rm e}^{-0.71\pm0.06}$
with a vertical scatter of 0.137 dex in $\log r_{\rm e}$.
This scatter translates to a 37 per cent error in distance per galaxy. 
The minimisation of $\log r_{\rm e}$ on the two photometric 
parameters gave the Photometric Plane
$r_{\rm e}\propto n^{0.86\pm0.13}\Sigma _{\rm e}^{-0.57\pm0.09}$, 
with a vertical scatter of 0.170 dex in $\log r_{\rm e}$, or a 48 per cent 
error in distance (see Figure~\ref{fig2}).

The above minimisation routine was tested on several galaxy sub-samples.
Removing the 8 Fornax elliptical galaxies and using only the Virgo
galaxies had no significant affect on the above results.  In using only
the elliptical galaxy sample from the Virgo and Fornax clusters (i.e.\ 
excluding the 11 S0 galaxies) also did not significantly alter the 
Fundamental Plane, 
but the Photometric Plane changed to 
$r_{\rm e}\propto n^{0.75\pm0.17}\Sigma _{\rm e}^{-0.75\pm0.12}$, 
with a vertical scatter of 0.153 dex in $\log r_{\rm e}$, 
or a 42 per cent distance error per galaxy. 
Use of the velocity dispersion data from McElroy (1995) 
was also explored and found not alter the above results beyond 
the 1$\sigma$ significance level. 

Using the complete sample of Virgo and Fornax galaxies and 
treating all variables equally (see, e.g., Feigelson \& Babu 1992), 
rather than minimising the residuals 
of just one variable (as done above), yielded the Photometric Plane 
$r_{\rm e}\propto n^{0.89\pm0.14}\Sigma _{\rm e}^{-0.60\pm0.09}$ 
and the Fundamental Plane 
$r_{\rm e}\propto \sigma _{0}^{1.22\pm0.11}\Sigma _{\rm e}^{-0.74\pm0.08}$. 
A consistent result was obtained after the exclusion of the S0
galaxies, and the separate exclusion of the Fornax galaxies. 
This Fundamental Plane is also, as one might expect, in complete 
agreement with the multivariate analysis performed by D'Onofrio et 
al.\ (1997), where they obtained 
$r_{\rm e}\propto \sigma _{0}^{1.26\pm0.09}\Sigma _{\rm e}^{-0.70\pm0.03}$
for their Virgo cluster data. 

Given the tight correlation between $n$ and $\sigma_0$, one
might not expect there to be any significant gains in using
all four parameters (i.e.\ $n, r_{\rm e}, <\mu>_{\rm e}$ and 
$\sigma_0$). 
Nonetheless, as the data is already at hand, a hyperplane 
was constructed and the scatter in 
$\log r_{\rm e}$ measured to be 0.125 dex (cf.\ 0.137 dex for the 
Fundamental Plane).  Removing the 8 Fornax
galaxies produced a scatter of 0.123 dex in $\log r_{\rm e}$, 
and the additional removal of the 11 S0 galaxies 
resulted in a scatter of only 0.105 dex
in $\log r_{\rm e}$ about the hyperplane.  This is equivalent to
an error of 27 per cent 
in distance per galaxy, whereas the Fundamental Plane gave 
a scatter of 0.125 dex (33 per cent error in distance) 
for this reduced galaxy sample, and the Photometric Plane had a 
scatter of 0.141 dex (38 per cent error in distance). 

The analysis above has been performed assuming that the Virgo and Fornax
clusters reside at the same distance from us. 
The latest results from the {\it HST Key project on the 
Extragalactic Distance Scale} (Freedman et al.\ 2001) find a 
Virgo-Fornax distance modulus of 0.51$\pm$0.21 mag. 
This result is based on Cepheid distances to 5 spiral galaxies in 
the Virgo cluster and 3 spiral galaxies in the Fornax cluster. 

We can quickly gauge the Photometric Plane's ability to 
estimate distances by adding some constant value to one clusters 
distance-dependent values of $\log r_{\rm e}$ (kpc) until the 
Photometric Plane of the combined Virgo+Fornax sample has a 
minimum scatter in $\log r_{\rm e}$. 
Both the E+S0 and the E-only galaxy samples were found to have a minimum
scatter when the Fornax galaxy radii were increased by
0.06 dex, giving a Virgo-Fornax distance modulus of +0.30 mag
and placing Fornax 15 per cent more distant than Virgo.  

A more accurate approach is to use the Working-Hotelling (1929) 
confidence bands described, and further developed, in Feigelson
and Babu (1992).  This technique has previously been applied 
to Fundamental Plane data in Graham (1998) to determine the 
intercept offset along the $\log r_e$ axis between the Virgo
and Fornax Fundamental Planes and is described there. 
The method allows for the fact that an error 
in slope to the fitted data will cause a greater discrepancy at
the ends of the relation, and hence the method gives more weight to
central data points. 
For our full galaxy sample, the Virgo and Fornax Fundamental Planes 
were found to have an offset of 0.114$\pm$0.067 dex 
in $\log r_{\rm e}$, implying a 
Virgo-Fornax distance modulus of 0.57$\pm$0.34 mag.  
The Photometric Planes had an offset of 0.117$\pm$0.088 dex in 
$\log r_{\rm e}$, implying a distance modulus of 0.59$\pm$0.44 mag.
The largish error can, in part, be attributed to the presence 
of only 8 galaxies in our Fornax cluster sample. 
Including those galaxies without velocity dispersion estimates 
but with measurements of $n$, the Virgo-Fornax distance modulus
derived from the Photometric Plane was found to be 0.62$\pm$0.30 
mag\footnote{The results given at the start of this section
did not change significantly when a Virgo-Fornax distance modulus 
of 0.6 was applied because there are only 8 Fornax galaxies.}.

%

\section{Discussion}

It had been our intention to analyze the Photometric Plane using 
those galaxies studied by Tonry et al.\ (1997) in their measurements 
of surface brightness fluctuations (SBF) for determining galaxy 
distances.
However, with regard to their data reduction process they write: 
``Finally, we make a mask of the obvious stars and companion galaxies 
in the cleaned image and determine the sky background by fitting 
the outer parts of the galaxy image with an $r^{1/4}$ profile plus 
sky level.''  Unfortunately, 
this procedure is fundamentally flawed because of its assumption 
that all elliptical galaxies have outer $r^{1/4}$ profiles.  Any real 
departures from an $r^{1/4}$ profile in the actual galaxy light profile 
will be largely erased by their tunable sky-level. 
This of course has many consequences: for example, 
colour gradients will be erroneous.  
%

This explains the strange behaviour of the $r^{1/4}$ models fitted 
to this data by Kelson et al.\ (2000a).  While their models match 
the outer light-profiles very well, they fail to 
the fit the inner regions of almost every galaxy.    
Although fitting an $r^{1/4}$ model to an $r^{1/n}$ profile may
not effect the $r_{\rm e}$ and $\Sigma_{\rm e}$ combination in 
the standard Fundamental Plane (Trujillo et al.\ 2001), the 
erroneous construction of an $r^{1/4}$ profile from an  
$r^{1/n}$ profile is another issue.  
This may possibly explain why the Fundamental Plane analysis 
of Kelson et al.\ (2000a) gave a notably different Hubble constant 
to the other 
techniques employed by the {\it HST Key Project on the Extragalactic 
Distance Scale} team\footnote{The use of aperture 
velocity dispersions within widely varying fractions of each 
galaxy's effective radius (as also used here) may have also been a 
contributing factor.}, and certainly rules out our hope to derive a 
Photometric Plane from these data.  

Across the Atlantic, 
D'Onofrio (2001a,b) recently performed a two-dimensional fit to 
the light distributions of the magnitude-limited galaxy 
samples discussed in section 2.  He fitted both seeing-convolved 
$r^{1/n}$ and seeing-convolved ($r^{1/n}$ + exponential) models.  
Using the parameters from his 
$r^{1/n}$ models fitted to the `genuine' elliptical galaxies, 
and the S\'ersic bulge parameters from the ($r^{1/n}$ + exponential) 
models fitted to the `genuine' S0 galaxies gave a Photometric 
(and Fundamental) Plane consistent (at the 1$\sigma$-level) with 
that derived above using the full galaxy sample of 38 objects 
mentioned in section 2. 
%
%
Using only the elliptical galaxies from D'Onofrio (2001) 
gave a level of scatter equivalent to that found previously; 
inclusion of the S0 galaxies resulted in an increased scatter. 
It should however be noted that the analysis by D'Onofrio (2001a,b) 
took into account several errors which can affect the structural 
parameters and in many ways this new data set is to be preferred.  
The planes constructed using both data sets are, reassuringly, the same. 


It is of course of interest to know what the prospects are for 
the Photometric Plane at higher redshifts.  Fortunately we have 
been able to immediately address this question using  
published S\'ersic model parameters from galaxies in the $z$=0.33 
cluster Cl 1358+62 (Kelson et al.\ 2000b, their table 1).
Kelson et al.\ fitted seeing-convolved S\'ersic models to their 
$z$=0.33 galaxy light profiles. 
Performing a regression that minimised the residuals in $\log r_{\rm e}$
for the combined E, E/S0 and S0 galaxy sample gave 
a scatter of 0.150 dex (an error of 41 per cent in 
distance\footnote{A distance error of 35 per cent was obtained 
when the later galaxy types used by Kelson et al.\ (2000b) were 
included here.}), the same as obtained 
above for the Virgo/Fornax elliptical galaxy sample. 
This is an extremely encouraging result, which could well 
be pursued with cluster samples spanning a range of redshifts
(e.g.\ Fasano et al.\ 2002).  

While the Photometric Plane used here can be viewed as a variant 
of the 
Fundamental Plane, in which the S\'ersic index $n$ has replaced 
the velocity dispersion, it can also be seen as an extension to 
the scalelength--shape\footnote{Given the results
of Figure~\ref{fig1}, the luminosity--shape ($L-n$) relation of 
Young \& Currie (1994) can be understood as a variant of the 
luminosity--velocity dispersion relation of Faber \& Jackson (1976).}
relation of Young \& Currie (1995; 2001).  
The scatter in $\log r_{\rm e}$ about the $\log r_{\rm e}$--$\log n$ 
relation for the present data sample is 0.35 dex, while the 
scatter we have computed about the Kormendy relation between 
$\log r_{\rm e}$ and $<\mu>_{\rm e}$ is 0.25 dex\footnote{These
values reduced to 0.28 dex and 0.21 dex when using only the Virgo
Elliptical galaxies.}.  
As we have seen above, using all three photometric parameters 
resulted in a tighter correlation. 

Concerns of parameter coupling in the fitting of the S\'ersic model,
and henceforth spurious correlations in our Photometric Plane, can
largely be laid to rest (see also Trujillo et al.\ 2001).  
We have used both the model-independent effective radius, and 
surface brightness at this radius, obtained directly from the image 
with no recourse to the S\'ersic model.   The S\'ersic index $n$ has 
of course come from the best-fitting $r^{1/n}$ model and this was 
used to convert $\mu_{\rm e}$ into $<\mu>_{\rm e}$.   However, 
the scatter about our Photometric Planes are the same no matter which 
of these two surface brightness terms we use.  

One big advantage in replacing the Fundamental Plane with a 
a relation based on purely photometric quantities is the 
considerable reduction in observational time and data analysis.
This has of course been recognised before, and led Scodeggio, 
Giovanelli \& Haynes (1997) to replace the stellar velocity 
dispersion term in the Fundamental Plane with the difference 
between the magnitude of a galaxy and that of the mode of the 
Gaussian luminosity function of the E/S0 galaxies.  This 
method could however only be applied to galaxy clusters whose
E/S0 luminosity function could be reliably determined.  
Additionally, the accuracy for distance determinations did not 
scale with the square root of the number of objects used to 
perform the fit.  Nonetheless, this was an interesting venture 
into the use of a purely photometric distance indicator for
early-type galaxies.


Assuming an isotropic velocity dispersion tensor, the absence of 
rotational energy, a constant mass-to-light ratio and a constant 
specific entropy for elliptical galaxies, 
Lima Neto et al.\ (1999) presented theoretical arguments for
the existence of what they termed an `Entropic Plane'.
This is a two-dimensional plane within the three-dimensional 
space of photometric parameters $\ln\Sigma_0$, $\ln h$ and $F(n)$, 
where $\Sigma_0$ is the central intensity of a bulge, $h$ is the radial
scalelength (rather than half-light radius), and $F(n)$ is a
function of the S\'ersic index $n$.  
 
Analyzing a sample of ordinary elliptical and dwarf spheroidal galaxies 
(with values of $n$ less than 4) taken from the rich clusters 
Coma and ABCG 85, and also from the group NGC~4839, 
Lima Neto et al.\ (1999) showed that 
the tilt to their observed Entropic planes did not quite match
the value expected from theory.  M\'arquez et al.\ (2000) 
subsequently identified one likely cause for the offset: 
Elliptical galaxies do not have a constant specific entropy but 
a value which increases with galaxy luminosity.  This conclusion was 
reached using the same data sample used in Lima Neto et al.\ (1999) 
and was further supported by simulations of hierarchial merging
galaxy formation.  One may well expect the level of entropy 
(disorder) to increase through mergers.  Nonetheless, despite this and 
other likely systematic differences with luminosity\footnote{Low
luminosity 
elliptical and S0 galaxies are known to contain significant 
rotation (e.g., Davies et al.\ 1983; Prugniel \& Simien 1994, 1996;  
Busarello et al.\ 1997; Graham et al.\ 1998), and dwarf elliptical 
galaxies are generally believed to be anisotropic (Bender, 
Burstein \& Faber 1992).  The issue of a constant M/L ratio 
with luminosity remains somewhat undecided, or at least has not 
yet been universally agreed upon (see, e.g., Prugniel \& Simien 1997).
}, 
a relatively tight plane was found to exist within a purely 
photometric set of parameters.  The physical grounds for this 
are pursued further in M\'arquez et al.\ (2001). 
Additional insight may come from the statistical mechanics of 
violent relaxation (Lynden-Bell 1967) which have been invoked to 
explain the range of elliptical galaxy light-profile shapes, such 
that $n$ increases as the luminosity (mass) does (Hjorth \& Madsen 1995).


To conclude, 
the Photometric Planes studied here display $\sim$15-30 per cent more 
scatter 
than the Fundamental Planes corresponding to the same galaxy sample. 
Due to the strong ($r_s$=0.82) correlation between the S\'ersic 
index $n$ and the 
central stellar velocity dispersion, hyperplanes which use all 
three photometric parameters plus the velocity dispersion have 
11-18 per cent less scatter than the Fundamental Plane. 
The scatter in $\log r_e$ about the Photometric Plane translates to 
a distance error of $\sim$38-48 per cent per galaxy.  A
scatter of 35-41 per cent was found for the $z$=0.33 cluster Cl 1358+62.  
The offset in $\log r_{\rm e}$ between the Virgo and Fornax 
Photometric Planes constructed using galaxies without velocity 
dispersion measurements gave a 
distance modulus of 0.62$\pm$0.30, implying the Fornax cluster is,
on average, 33$\pm$15 per cent more distant.  This result is in good 
agreement with other accurate distance determinators, and indicates
the applicability of the Photometric Plane to practical situations. 
Although the Fundamental Plane appears to have less scatter than 
the Photometric Plane, we would recommend authors construct the 
Photometric Plane and check if its level of accuracy is sufficient
for their needs before pursuing additional (expensive) kinematical 
data.


\label{lastpage}


\begin{thebibliography}{99999}
\bibitem{Bet92}Bender R., Burstein D., Faber S.M., 1992, ApJ, 399, 462 
\bibitem{Bet97}Busarello G., Capaccioli M., Capozziello S., Longo G., Puddu E., 1997, A\&A, 320, 415
\bibitem{CCD93}Caon N., Capaccioli M., D'Onofrio M., 1993, MNRAS, 265, 1013
\bibitem{CCD94}Caon N., Capaccioli M., D'Onofrio M., 1994a, A\&AS, 106, 199 
\bibitem{Cet94}Caon N., Capaccioli M., D'Onofrio M., Longo G., 1994b, A\&A, 286, L39
\bibitem{CCR90}Caon N., Capaccioli M., Rampazzo R., 1990, A\&AS, 86, 429 
\bibitem{Det83}Davies R.L., Efstathiou G., Fall S.M., Illingworth G., Schecter P.L., 1983, ApJ, 266, 41
\bibitem{DaD87}Djorgovski S., Davis M., 1987, ApJ, 313, 59
\bibitem{Don1a}D'Onofrio, 2001a, MNRAS, 326, 1508
\bibitem{Don1b}D'Onofrio, 2001b, MNRAS, 326, 1517
\bibitem{DCC94}D'Onofrio M., Capaccioli M., Caon N., 1994, MNRAS, 271, 523
\bibitem{Det97}D'Onofrio M., Capaccioli M., Zaggia S. R., Caon N. 1997, MNRAS, 289, 847
\bibitem{Det87} Dressler A., Lynden-Bell D., Burstein D., Davies R.L., Faber S.M., Terlevich R.J., Wegner G., 1987, ApJ, 313, 42
\bibitem{Fet97}Faber S.M., et al., 1997, AJ, 114, 1771
\bibitem{FaJ76}Faber S.M., Jackson R.E., 1976, ApJ, 204, 668 
\bibitem{Fas02}Fasano G., Bettoni D., D'Onofrio M., Kjaergaard P., Moles M., A\&A, in press (astro-ph/0203013)
\bibitem{FaB92}Feigelson E.D., Babu G.J., 1992, ApJ, 397, 55
\bibitem{Fet94}Ferrarese L., van den Bosch F.C., Ford H.C., Jaffe W., O'Connell R.W., 1994, AJ, 108, 1598
\bibitem{Fet01}Freedman W., et al., 2001, ApJ, 553, 47
\bibitem{Gra98}Graham A.W., 1998, MNRAS, 295, 933
\bibitem{GaC97}Graham A.W., Colless M.M., 1997, MNRAS, 287, 221 
\bibitem{Grt98}Graham A.W., Colless M.M., Busarello G., Zaggia S., Longo G., 1998, A\&AS, 133, 325
\bibitem{Get01}Graham A.W., Erwin P., Trujillo I., Caon N., 2001, ApJ, 563, L11
\bibitem{GTC01}Graham A.W., Trujillo I., Caon N., 2001, AJ, 122, 1707
\bibitem{HaM95}Hjorth J., Madsen J., 1995, ApJ, 445, 55
\bibitem{JaB97}Jerjen H., Binggeli B., 1997, in The Nature of Elliptical Galaxies; The Second Stromlo Symposium, ASP Conf.\ Ser., 116, 239
\bibitem{Ke00a}Kelson D.D., et al., 2000a, ApJ, 529, 768
\bibitem{Ke00b}Kelson D.D., Illingworth G.D., van Dokkum P.G., Marijn, F., 2000b, ApJ, 531, 137
\bibitem{Kho0a}Khosroshahi H.G., Wadadekar Y., Kembhavi A., Mobasher B., 2000a, ApJ, 531, L103
\bibitem{Kho0b}Khosroshahi H.G., Wadadekar Y., Kembhavi A., 2000b, ApJ, 533, 162
\bibitem{Let95}Lauer T.R., et al., 1995, AJ, 110, 2622
\bibitem{LGM99}Lima Neto G.B., Gerbal D., M\'arquez I., 1999, MNRAS, 309, 481
\bibitem{Lyn67}Lynden-Bell D., 1967, MNRAS, 136, 101
\bibitem{Met00}M\'arquez I., Lima Neto G.B., Capelato H., Durret F., Gerbal D., 2000, A\&A, 353, 873 
\bibitem{Met01}M\'arquez I., Lima Neto G.B., Capelato H., Durret F., Lanzoni B., Gerbal D., 2001, A\&A, 379, 767 
\bibitem{McE95}McElroy D.B., 1995, ApJS, 100, 105
\bibitem{MaH01}M\"ollenhoff C., Heidt J., 2001, A\&A, 368, 16 
\bibitem{PaS94}Prugniel Ph., Simien F., 1994, A\&A 282, L1 
\bibitem{PaS96}Prugniel Ph., Simien F., 1996, A\&A 309, 749 
\bibitem{PaS97}Prugniel Ph., Simien F., 1997, A\&A, 321, 111
\bibitem{SGH97}Scodeggio M., Giovanelli R., Haynes M.P., 1997, AJ, 113, 2087
\bibitem{Set68}S\'ersic J.-L., 1968, Atlas de Galaxias Australes (Cordoba: Observatorio Astronomico)
\bibitem{Tet97}Tonry J., Blakeslee J.P., Ajhar E.A., Dressler A., 1997, ApJ, 475, 399
\bibitem{TGC01}Trujillo I., Graham A.W., Caon N., 2001, MNRAS, 326, 869
\bibitem{WaH29}Working H., Hotelling H., 1929, J. Am. Stat. Assoc. Suppl., 24 73
\bibitem{YaC94}Young C.K., Currie M.J., 1994, MNRAS, 268, L11
\bibitem{YaC95}Young C.K., Currie M.J., 1995, MNRAS, 273, 1141
\bibitem{YaC01}Young C.K., Currie M.J., 2001, A\&A, 369, 736
\end{thebibliography}
\end{document}